\documentclass[aps, prl, twocolumn, amssymb, amsmath, superscriptaddress]{revtex4-1}

\usepackage{bm}
\usepackage{times}
\usepackage{makecell}
\usepackage{graphicx}
\usepackage[colorlinks,citecolor=blue,linkcolor=blue]{hyperref}
\usepackage{array}
\usepackage{float}
\usepackage{colortbl}
\usepackage{caption}
\usepackage{multirow}
\usepackage{tabularx}

\usepackage{amsmath}

\usepackage{gensymb}
\graphicspath{ {./} }

\begin{document}
\title{Superconductivity and unexpected chemistry of germanium hydrides under pressure}

\author{M. Mahdi Davari Esfahani}
\affiliation{Department of Geosciences, Center for Materials by Design, and Institute for Advanced Computational Science,
State University of New York, Stony Brook, NY 11794-2100, USA}
\author{Artem R. Oganov}\email{artem.oganov@stonybrook.edu}
\affiliation{Department of Geosciences, Center for Materials by Design, and Institute for Advanced Computational Science,
State University of New York, Stony Brook, NY 11794-2100, USA}
\affiliation{Skolkovo Institute of Science and Technology, Skolkovo Innovation Center, 3 Nobel St., Moscow 143026, Russia}
\affiliation{Department of Problems of Physics and Energetics, Moscow Institute of Physics and Technology, 9 Institutskiy Lane,
Dolgoprudny City, Moscow Region 141700, Russia}
\affiliation{International Center for Materials Design, Northwestern Polytechnical University, Xi'an,710072, China}
\author{Haiyang Niu}
\affiliation{Department of Geosciences, Center for Materials by Design, and Institute for Advanced Computational Science,
State University of New York, Stony Brook, NY 11794-2100, USA}
\author{Jin Zhang}
\affiliation{Department of Geosciences, Center for Materials by Design, and Institute for Advanced Computational Science,
State University of New York, Stony Brook, NY 11794-2100, USA}

\date{\today}

\begin{abstract}
Following the idea that hydrogen-rich compounds might be high-{\it T\/}$_c$ superconductors at high pressures, and the very recent breakthrough in predicting and synthesizing hydrogen sulfide with record-high {\it T\/}$_c$ = 203 K, ab initio evolutionary algorithm for crystal structure prediction was employed to find stable germanium hydrides. 
In addition to the earlier structure of germane with space group {\it Ama2\/}, we propose a 
new {\it C2/m\/} structure, which is energetically more favorable at pressures above 278 GPa (with inclusion of zero point energy).
Our calculations indicate metallicity of the new {\it C2/m\/} phase of germane with {\it T\/}$_c$ = 67 K at 280 GPa. 
Germane is found to exhibit thermodynamic instability to decomposition to hydrogen and the new compound Ge$_3$H$_{11}$ at pressures above 300 GPa. Ge$_3$H$_{11}$ with space group {\it I$\bar{4}$m2\/} is found to become stable at above 285 GPa with  {\it T\/}$_c$ = 43 K.
We find that the pressure-induced phase stability of germanium hydrides is distinct from its analogous isoelectronic systems, e.g., Si-hydrides and Sn-hydrides. 
Superconductivity stems from large electron-phonon coupling associated with the wagging, bending and stretching intermediate-frequency modes derived mainly from hydrogen.

\end{abstract}


\flushbottom
\maketitle

% * <MAHDI.DAVARI@STONYBROOK.EDU> 2015-02-09T12:07:31.197Z:
\thispagestyle{empty}

%\section*{Introduction}
High-throughput materials discovery using first-principles density functional theory (DFT) \cite{kohn1965self} has motivated many experimental studies. For years, scientists have been trying to find the best way to design high-temperature superconductors. 
It has been confirmed that high-{\it T\/}$_c$ superconductivity can be found in systems with light elements. Hydrogen is the lightest element with rich structures and properties under high pressures. Within BCS (Bardeen-Cooper-Schrieffer) theory of superconductivity \cite{bardeen1957theory}, high vibrational frequencies of hydrogen atoms and often high electron-phonon coupling make it possible to expect high-{\it T\/}$_c$ in metallic hydrogen and hydrogen-rich hydrides.

However, metallic hydrogen seems to require very high pressure $\sim$ 400 GPa and proved elusive. 
Therefore, chemical precompression by alloying with heavy element was proposed \cite{Ashcroft2004}. 
Many theoretical and experimental studies have been motivated by this idea
 to seek and design new high-{\it T\/}$_c$ superconductors at high pressures \cite{PhysRevLett.101.107002,zhong2016tellurium,errea2016quantum,zhong2012structural,PhysRevB.84.054543,C4RA13183E,PhysRevB.91.180502,struzhkin2015superconductivity,Eremets1506,drozdov2015superconductivity,drozdov2015conventional}.

In a recent breakthrough discovery, which was first predicted by the evolutionary algorithm USPEX coupled with DFT studies \cite{duan2014pressure}, high-temperature superconductivity with a transition temperature ({\it T\/}$_c$) of 203 K in hydrogen sulfide H$_3$S under pressure 200 GPa has been reported by Drozdov {\it et al.\/} \cite{drozdov2015conventional}. This discovery not only set a record high-{\it T\/}$_c$ for a conventional phonon-mediated mechanism, but also raised hopes of reaching room-temperature superconductivity in hydrogen-rich metallic alloys. This realization is the best argument to show the predictive power of DFT-based structure prediction and electron-phonon coupling calculations, and opens up avenues for discovering superconductors based on this approach.

Successful synthesis of hydrogen sulfides with superconducting properties was followed by a second high-{\it T\/}$_c$ hydrogen-rich compound at high pressure (PH$_3$) synthesized by Drozdov {\it et al.\/}\cite{drozdov2015superconductivity}. 
Prior to H$_3$S, the highest experimentally observed {\it T\/}$_c$ in conventional superconductors, which obey the BCS theory was in MgB$_2$, which opened avenues for searching for higher {\it T\/}$_c$ superconductors. However, other magnesium borides Mg$_x$B$_y$ were shown to exhibit poor superconductivity with {\it T\/}$_c$ $<$ 3 K. 
Besides these efforts, other superconductors have been predicted in hydrogen-rich compounds.
In group-IV hydrides, SiH$_4$ has been predicted to have {\it T\/}$_c$ = 20-75 K \cite{PhysRevLett.101.077002}, while experiment got a lower value of 17 K \cite{Eremets1506}. 
Disilane (SiH$_8$) has been predicted to favor {\it Ccca\/} structure with {\it T\/}$_c$ of 98-107 K at 250 GPa \cite{strobel2009novel}. 
Our work on tin hydrides showed rich chemistry of that system with high-{\it T\/}$_c$ superconductivity. 
Tin hydrides have been predicted to form at high pressures, exhibiting high {\it T\/}$_c$ of 81, 93, 97 K for SnH$_8$, SnH$_{12}$ and SnH$_{14}$ at 220, 250 and 300 GPa, respectively \cite{esfahani2016superconductivity}.
In addition, novel linear and bent formation of H$_3$ and H$_4$ have been predicted to form in high-pressure phases of SnH$_8$, SnH$_{12}$ and SnH$_{14}$ \cite{esfahani2016superconductivity}.

Germanium (Ge) is in the same group-IV and is isovalent to Sn. One can expect germanium to exhibit similar chemistry as tin, but its smaller atomic radius and slightly higher electronegativity than Sn result in quite a different chemistry.

Germane (GeH$_4$) phases have been explored by Gao {\it et al.\/} \cite{PhysRevLett.101.107002} and their results show {\it C2/c\/}-GeH$_4$ becomes stable at pressures above 196 GPa (including zero point energy (ZPE)) against decomposition into H and Ge. However, stability against  decomposition into the elements is not a particularly stringent test, and stability against separation into other phases, e.g., GeH$_4$ into Ge$_2$H and H$_2$, which is important for understanding decomposition mechanism, should be taken into account. {\it C2/c\/} was predicted to be a superconductor with {\it T\/}$_c$ = 64 K at 220 GPa. In a recent theoretical study, a more energetically stable structure of germane (with symmetry group {\it Ama2\/}) was predicted by Zhang {\it et al.\/} to have {\it T\/}$_c$ of 47-57 K \cite{C5CP03807C}. Now, with major progress of computational methods (enabling, for example, variable-composition searches), we can address all the outstanding issues.

We systematically explored the high-pressure phase diagram of Ge-H system with using evolutionary variable-composition search implemented in the USPEX code \cite{Oganov2011,Oganov01012010,Glass2006,Lyakhov2010} from ambient pressure to 400 GPa. 
The effectiveness of this method has been shown by the prediction of high-pressure structures of various systems that were subsequently confirmed experimentally (e.g., \cite{Zhang20122013,Oganov2009,mannix2015synthesis}).
In this method, we created initial generation of structures and compositions using the random symmetric algorithm \cite{lyakhov2013new}. 
Subsequent generations were obtained using heredity, transmutation, softmutation, and random symmetric generator \cite{lyakhov2013new}.
Ge hydrides, in comparison with other hydrides of the same group, e.g., Si \cite{strobel2009novel,PhysRevLett.101.077002}, Sn \cite{esfahani2016superconductivity} which often show simpler phase diagram, exhibit a unique and complex potentianl energy landscape. Unexpected stoichiometries Ge$_3$H, Ge$_2$H, GeH$_3$, GeH$_4$ and Ge$_3$H$_{11}$ emerge as stable  at megabar pressures.

The underlying structure relaxations were carried out using VASP package \cite{PhysRevB.54.11169} in the framework of density functional theory (DFT) and using PBE-GGA (Perdew-Burke-Ernzerhof generalized gradient approximation) \cite{PhysRevLett.77.3865}. 
The projector-augmented wave approach (PAW) \cite{PhysRevB.50.17953} was used to describe the core electrons and their effects on valence orbitals. A plane wave kinetic energy cutoff of 1000 eV for hard PAW potentials and dense Monkhorst-Pack {\it k\/}-points grids with reciprocal space resolution 2$\pi$ $\times$ 0.03 \AA$^{-1}$ were employed \cite{PhysRevB.13.5188} to sample the Brillouin zone.

Phonon frequencies and superconducting properties were calculated using density functional perturbation theory as implemented in {\sc Quantum ESPRESSO} package \cite{QE2009}. 
PBE-GGA functional is used for this part. A plane-wave basis set with a cutoff of 80 Ry gave a convergence in energy with a precision of 1 meV/atom. We used valence electron configurations of 3d$^{10}$ 4s$^2$ 4p$^2$  and 1s$^1$ for germanium and hydrogen, respectively.
Thermodynamic properties of germanium hydrides were calculated using the PHONOPY package with the implemented frozen-phonon approach \cite{phonopy}.

The electron-phonon coupling (EPC) parameter $\lambda$ was calculated 
using 5$\times$5$\times$2 and 4$\times$4$\times$4 {\it q\/}-point meshes for {\it I$\bar{4}$m2\/}-Ge$_3$H$_{11}$ and {\it C2/m\/}-GeH$_4$, respectively. Denser {\it k\/}-point meshes, 20$\times$20$\times$8 and 16$\times$16$\times$16 
were used in the calculations of the electron-phonon interaction matrix elements.
The superconducting {\it T\/}$_c$, was estimated using the Allen-Dynes modified McMillan equation \cite{PhysRevB.12.905}.%:

 The energetic stability of a variety of Ge$_x$H$_y$ (x + y $<$ 20) compounds was evaluated using the thermodynamic convex hull construction at different pressures (see Supplemental Material Fig. S1). %, as depicted in Fig. 1.
 To our surprise, in addition to reproducing various structures of Ge-H system \cite{C4RA13183E,PhysRevLett.101.107002,C5CP03807C,0295-5075-90-6-66006}, Ge \cite{PhysRevB.62.11388} and H$_2$ \cite{Pickard2007}, previously unreported and unexpected composition of germanium hydrides Ge$_3$H$_{11}$ was found to be stable in wide pressure ranges.

Below 200 GPa, no hydrogen-rich composition is stable against decomposition into the elements. It is consistent with not having any solid H-rich Ge-hydrides at low pressures, although using {\it in situ\/} gas-condensation techniques, Maley {\it et al.\/} showed germane can form at ambient pressure \cite{maley2008solid}. 
 Increasing pressure decreases formation enthalpies, implying a tendency for Ge-hydrides to be stabilized under further compression.  
Phases of elemental H and Ge for the convex hull construction were obtained from structure  search, in good agreement with the ones reported in Ref \cite{Pickard2007} for hydrogen and for elemental Ge, we obtained a complex phase diagram with at least four phase transitions between 70 and 400 GPa, which are in good agreement with Ref \cite{PhysRevB.62.11388}.

\begin{figure*}[h]
\centerline{\includegraphics[width=1.0\textwidth]{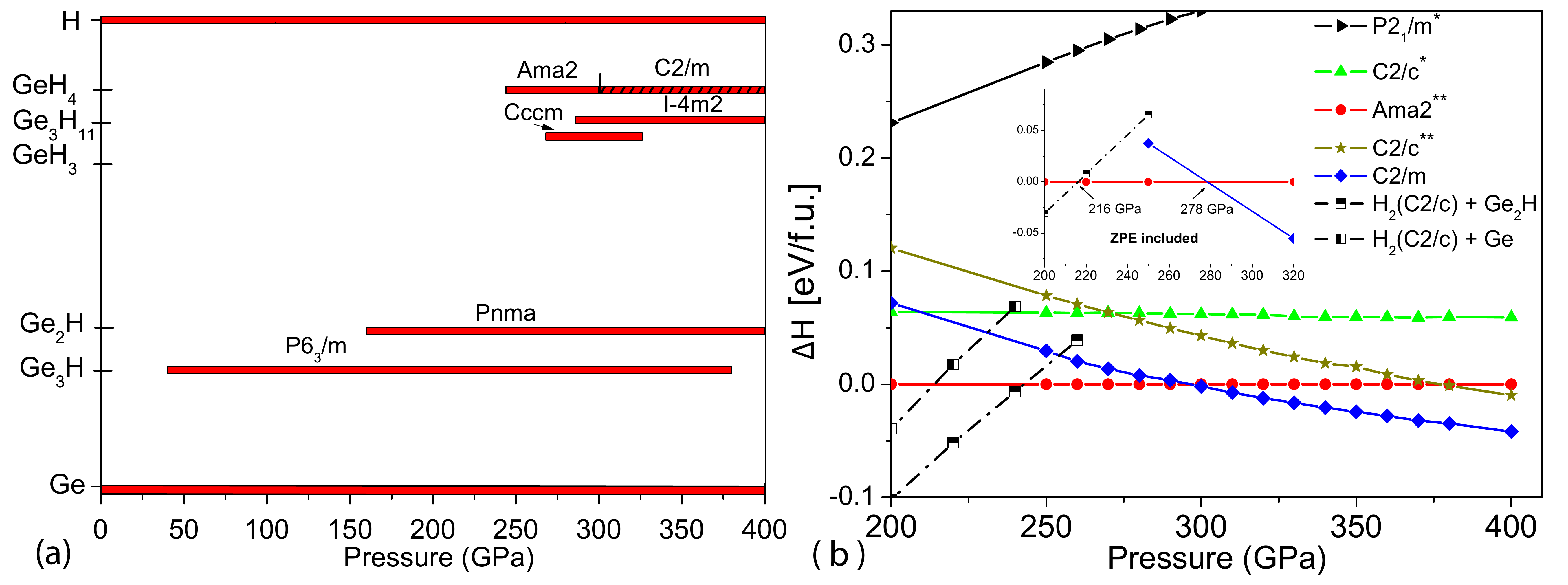}}
\caption{(a) Predicted pressure-composition phase diagram of the Ge-H system. The dashed areas represent thermodynamically metastable structures. (b) The enthalpies per formula unit of various structures of germane as a function of pressure with respect to the previously reported {\it Ama2\/} structure \cite{C5CP03807C}. Decomposition (GeH$_4$) enthalpies are calculated by adopting the {\it C2/c\/} structure for H$_2$ (Ref. \cite{Pickard2007}) and Ge$_2$H in the {\it Pnma\/} structure. 
The elemental decomposition enthalpies are also added for comparison. Inset: Enthalpies for {\it C2/m\/} structure relative to {\it Ama2\/} structure with the zero-point corrections.
The superscript "$^*$" and "$^{**}$" represent the structures predicted by Gao {\it et al.\/} \cite{PhysRevLett.101.107002} and Zhang {\it et al.\/} \cite{C5CP03807C}, respectively.}
\label{fig2}
\end{figure*}

At 250 GPa, the tetragonal Ge$_3$H$_{11}$ with space group {\it I$\bar{4}$m2\/} is still metastable and lies just above the tie-line joining {\it Ama2\/}-GeH$_4$ and {\it Pnma\/}-Ge$_2$H. At 300 GPa, we predict stable phases: Ge$_3$H ({\it P6$_3$/m\/}), Ge$_2$H ({\it Pnma\/}) and GeH$_3$ ({\it Cccm\/}) in accord with previous predictions \cite{C4RA13183E,C5CP03807C}. In addition, we also found unexpected composition Ge$_3$H$_{11}$ that appears in the H-rich region, its structure featuring GeH$_{12}$ distorted icosahedra and GeH$_{16}$ Frank-Casper polyhedra. 
Moreover, germane transforms to a new monoclinic phase with space group {\it C2/m\/} with 3 f.u./cell at above 300 GPa (278 GPa with inclusion of ZPE), which is lower in enthalpy than all previously proposed structures \cite{PhysRevLett.101.107002,C4RA13183E,C5CP03807C} (see also Fig. \ref{fig2}(b)).

The stability fields of solids Ge$_3$H, Ge$_2$H, GeH$_3$, Ge$_3$H$_{11}$ and GeH$_4$ are illustrated in pressure-composition phase diagram of the Ge-H system, as shown in Fig. \ref{fig2}(a).
 Ge-rich compounds tend to stabilize at lower pressure ($<$200 GPa), while higher pressure ($>$200 GPa) is required for H-rich compounds to form. To the best of our knowledge, these unexpected, yet complex stoichiometries have not been reported in group-IV hydrides except MH$_4$ (M = Si, Sn, Pb). This rich chemistry makes Ge hydrides of special interests. It can be seen that the formation of Ge$_3$H$_{11}$ at 285 GPa lowers the convex hull and finally around 300 GPa causes GeH$_4$ to become thermodynamically metastable. The dynamical stability of structures shown in Fig. \ref{fig2}(a). were confirmed in their pressure ranges of stability via phonon calculations.

\begin{figure*}[h]
\centerline{\includegraphics[width=1.0\textwidth]{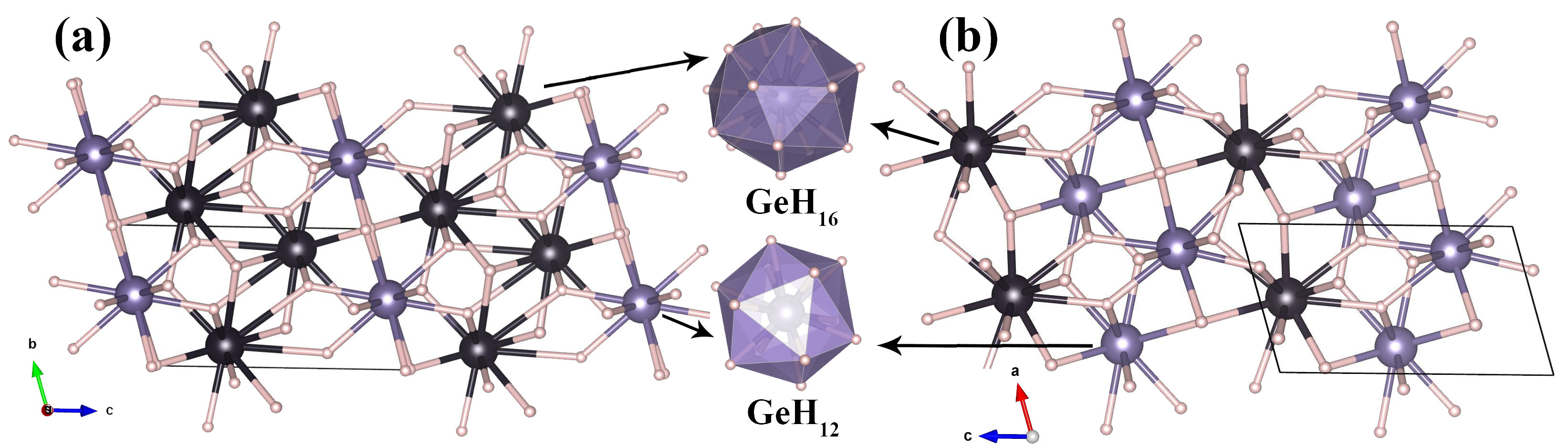}}
\caption{Predicted structures of Ge-H compounds at high pressures: (a) GeH$_4$ in the {\it C2/m\/} structure, (b) Ge$_3$H$_{11}$ in the {\it I$\bar{4}$m2\/} structure. Small and large spheres represent H and Ge atoms, respectively. Different color of germanium atoms represent diffferent type of polyhedra, i.e., black spheres represent GeH$_{16}$ polyhedra and purple spheres show GeH$_{12}$ icosahedra.} 
\label{fig3}
\end{figure*}

GeH$_4$ was predicted to become stable against decomposition into the elements at above 225 GPa (196 GPa with the inclusion of zero-point energy) \cite{gao2008superconducting}, while our results reveal lower enthalpy of Ge$_2$H+H$_2$ indicating the need for somewhat higher pressure 244 GPa (216 GPa with ZPE inclusion) for GeH$_4$ to be stabilized (see Fig. \ref{fig2}(b)-inset). Upon increasing pressure, the {\it Ama2\/} structure of GeH$_4$ transforms into the {\it C2/m\/} structure at 300 GPa. Structures predicted in the literature are also included for comparison.  In the {\it Ama2\/} $\rightarrow$ {\it C2/m\/} transition, the coordination number of Ge atoms increases from 10 to 12 and 16 with the formation of GeH$_{12}$ distorted icosahedra and GeH$_{16}$ Frank-Casper polyhedra at 300 GPa (Fig. \ref{fig3}(a)-inset). 
In addition, the average Ge-H bond lengths slightly increase from 1.698 \AA\ to 1.704 \AA\ in the {\it Ama2\/} $\rightarrow$ {\it C2/m\/} transition.

GeH$_4$ is unstable against decomposition to H$_2$ ({\it Cmca\/}) and Ge$_3$H$_{11}$ ({\it I$\bar{4}$m2\/}) at pressures above 300 GPa, according to the convex hull (see Supplemental Material Fig. S1(b) and (c)). Similarly, GeH$_3$ decomposes to Ge$_2$H and Ge$_3$H$_{11}$.

%#### 
Both in %The arrangement of germanium atoms in
 {\it I$\bar{4}$m2\/}-Ge$_3$H$_{11}$ and {\it C2/m\/}-GeH$_4$, %is relatively similar, in which, 
 each Ge atom is coordinated with 12 and 16 H atoms making distorted icosahedra and GeH$_{16}$  Frank-Casper polyhedra (see Fig. \ref{fig3}). The average Ge-H bond lengths are 1.660 and 1.704 \AA\ in {\it I$\bar{4}$m2\/}-Ge$_3$H$_{11}$ and {\it C2/m\/}-GeH$_4$ at 300 GPa, respectively.    
Unlike other compressed hydrides \cite{zhong2016tellurium,esfahani2016superconductivity,duan2015decomposition,hooper2013polyhydrides}, there are no bonds between H atoms.

%###
As shown in the Fig. \ref{fig3}., liberating one hydrogen atom from a 3 f.u. cell turns a GeH$_{16}$ polyhedra to a less coordinated germanium atom and leads to the formation of a distorted icosahedron, i.e., GeH$_4$ consists of two GeH$_{16}$ polyhedra and a GeH$_{12}$ icosahedron, however, Ge$_3$H$_{11}$ turns out to have one GeH$_{16}$ polyhedron and two distorted icosahedra. The detailed crystallographic data are listed in Table I.

%###ZPE
Because of high concentration of hydrogen in GeH$_4$, contribution of zero-point energy (ZPE) would be important in determining the relative stability of hydrogen-rich phases \cite{jacs.5b10180,wang2012superconductive,gao2008superconducting,errea2016quantum,esfahani2016superconductivity}. However, our results show that ZPE does not change the topology of the phase diagram of GeH$_4$, and quantitative effects are just moderate shifts in transition pressures. For example, the inclusion of ZPE lowers the formation enthalpies of {\it Ama2\/} and {\it C2/m\/} structures and shifts the transition pressure {\it Ama2\/} $\rightarrow$ {\it C2/m\/} from 300 to 278 GPa indicating enhanced stability of {\it C2/m\/} phase owing to ZPE (see Fig. \ref{fig2}(b)-inset).

%####
Analyzing the electronic band structures of GeH$_4$ ({\it C2/m\/}) and Ge$_3$H$_{11}$ ({\it I$\bar{4}$m2\/}) (see Supplemental Material Fig. S2(a) and (b)) indicates indirect band overlap which results in metallic behavior, with highly dispersive bands crossing the Fermi level, these bands being
basically due to germanium states with \textit{p}-character and marginally due to hydrogen states with \textit{s}-character. These H-derived states near Fermi level resemble those of solid metallic hydrogen. 
The {\it C2/m\/} structure is a metal with parabolic dispersive bands crossing the Fermi level along the A-$\Gamma$-Z symmetric line with several electron and hole pockets at the Fermi level. 
In the energy region near E$_f$, the DOS of Ge is about two times that of H, which indicates the dominance of Ge atoms contribution in the bands near the Fermi level. 
The total DOS at E$_f$, N(E$_f$), is 0.27 states/eV/f.u. for the {\it C2/m\/}-GeH$_4$ structure at 300 GPa, while we see higher N(E$_f$) = 0.31 for {\it Ama2\/} phase at 300 GPa. 
The Fermi levels of GeH$_4$ and Ge$_3$H$_{11}$ fall %in a shallow minimum%
on a shoulder of the density of states, while the record {\it T\/}$_c$ in H$_3$S is explained to be due to the van Hove singularity close to the Fermi level \cite{PhysRevB.93.104526,PhysRevB.94.064507}, therefore, doping can be expected to raise N(E$_f$) and {\it T\/}$_c$ values. 
These values of DOS at the Fermi level N(E$_f$) are lower than those in H$_3$S (0.54 states/eV/f.u.). 

%#####ELFFFF

%####PHONON

\begin{figure*}[h]
\centerline{\includegraphics[width=\textwidth]{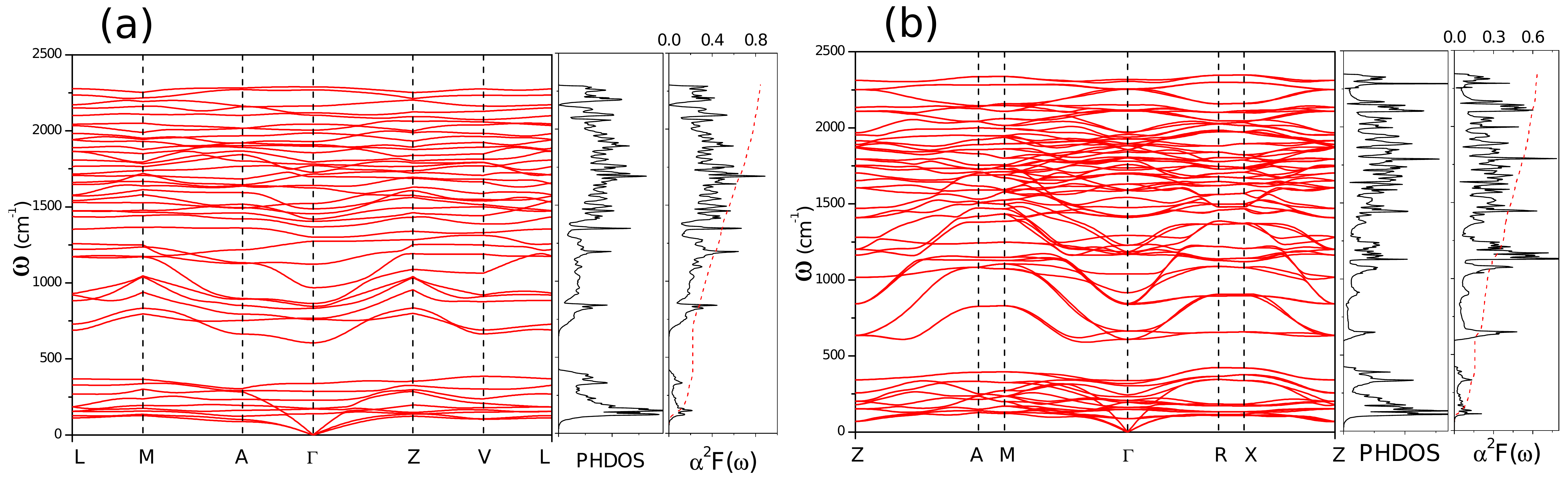}}
\caption{Calculated phonon dispersion curves, phonon density of states (PHDOS), Eliashberg EPC spectral functions $\alpha^2$F($\omega$) and electron-phonon integral $\lambda$($\omega$) of (a) GeH$_4$ [{\it C2/m\/}] at 300 GPa, (b) Ge$_3$H$_{11}$ [{\it I$\bar{4}$m2\/}] at 300 GPa.} 
\label{fig5}
\end{figure*}

To probe the possible superconducting behavior, electron-phonon coupling (EPC) calculations were performed for {\it C2/m\/}-GeH$_4$ and {\it I$\bar{4}$m2\/}-Ge$_3$H$_{11}$ structures at 280, 300 and 320 GPa. Phonon dispersions, phonon density of states, the corresponding Eliashberg spectral function $\alpha^2F$($\omega$)  and the EPC parameter $\lambda$ as a function of frequency are calculated and shown in Fig. \ref{fig5}(a) and (b) for {\it C2/m\/}-GeH$_4$ and {\it I$\bar{4}$m2\/}-Ge$_3$H$_{11}$ at 300 GPa, respectively.

The low frequency bands below 430 cm$^{-1}$ are mainly from the strongly coupled vibrations between Ge and H that contribute about 26\%  (25\%) of the total $\lambda$, while higher-frequency phonons, predominantly wagging, bending and stretching modes between 550-2300 cm$^{-1} $ are mostly related to the H atoms bonded to Ge and contribute 74\%  (75\%) of $\lambda$ of {\it C2/m\/}-GeH$_4$ ({\it I$\bar{4}$m2\/}-Ge$_3$H$_{11}$) phase.

The resulting integral $\lambda$ and logarithmic average phonon frequencies ($\omega_{log}$) are calculated using the Eliashberg formalism and then {\it T\/}$_c$ values are estimated using Allen-Dynes modified McMillan equation with using Coulomb pseudopotential parameters $\mu^*$ = 0.1 and 0.13 as commonly accepted values. Table II summarizes data for the total EPC parameters $\lambda$, logarithmic phonon average frequencies and corresponding {\it T\/}$_c$ values at given pressures.

Hard phonons in H-rich materials are described to play an important role in high-{\it T\/}$_c$ superconductivity \cite{PhysRevLett.21.1748}, but because such hard phonons do not always produce large coupling constant, high-{\it T\/}$_c$ superconductivity is still elusive. 
In the {\it C2/m\/}-GeH$_4$ structure, high-frequency vibrations that contribute the most to EPC parameter, produce a larger coupling constant i.e., 25\% higher than similar frequency modes in {\it I$\bar{4}$m2\/}-Ge$_3$H$_{11}$.  Additional flat bands in the high frequency region of {\it C2/m\/}-GeH$_4$ phonon modes, can be ascribed to the higher coupling constant and eventually result in getting higher {\it T\/}$_c$ value for {\it C2/m\/}-GeH$_4$.

%######
We investigated the pressure dependence of the critical transition temperature. %. The results depicted in Fig. 6, 
 The results show that the calculated {\it T\/}$_c$ decreases monotonically with pressure
with approximate rates of -0.19 and -0.20 K/GPa for {\it C2/m\/}-GeH$_4$ and {\it I$\bar{4}$m2\/}-Ge$_3$H$_{11}$  %(67.3 K at 280 GPa and 59.8 K at 320 GPa) 
in the pressure range 280-320 GPa.
Higher {\it T\/}$_c$ of the {\it C2/m\/} phase, compared to the previously reported phase  {\it Ama2\/}-GeH$_4$, can be  related to the considerably higher average phonon frequency, which also can be explained through a BCS mechanism.

%\section{Conclusions}

In summary, we explored high-pressure phase diagram of the Ge-H binary system by exploring its compositional and configurational space with an evolutionary crystal structure prediction method. Based on analysis of current and prior theoretical studies on Ge-hydrides, we have established 
thermodynamically stable phases, superconducting properties, structural features and new decomposition lines in the pressure range 0-400 GPa.

At 250 GPa, all the stoichiometries Ge$_2$H, Ge$_3$H and GeH$_4$ are energetically stable against any decomposition into the elements or any other compounds. At 300 GPa, GeH$_3$ and Ge$_3$H$_{11}$ become stable, while GeH$_4$ becomes unstable.

A unique metallic phase of germane with {\it C2/m\/} space group is found to be energetically more favorable than all previously proposed structures at pressures above 278 GPa (if zero-point energy is included). Our results reveal that germane decomposes to hydrogen and the newly found compound Ge$_3$H$_{11}$ at the pressures above 300 GPa. 
According to electron-phonon coupling calculations, {\it C2/m\/}-GeH$_4$ and {\it I$\bar{4}$m2\/}-Ge$_3$H$_{11}$ are excellent superconductors with high-{\it T\/}$_c$ of 67 K and 43 K for {\it C2/m\/}-GeH$_4$ at 280 GPa and {\it I$\bar{4}$m2\/}-Ge$_3$H$_{11}$ at 285 GPa, respectively.

\section{Acknowledgements}
We thank DARPA (grant W31P4Q1210008) and the Russian Science Foundation (grant 16-13-10459) for financial support. Calculations were performed on XSEDE facilities and on the cluster of the Center for Functional Nanomaterials, Brookhaven National Laboratory.

\bibliography{scibib}{}
\bibliographystyle{apsrev4-1}

\newpage
\clearpage

\begin{table}[ht]
\centering	
\caption{\bf Predicted crystal structures of Ge$_3$H$_{11}$ and GeH$_4$ at 300 GPa.}
\begin{tabular*}{9cm}{@{\extracolsep{\fill}}cccccc}%{c c c c c c} 
 \hline\hline
 Phase & Lattice & Atom & x & y & z \\ [0.5ex] 
             &parameters &   &   &   \\ [0.5ex] 
 \hline
 {\it I$\bar{4}$m2\/}-Ge$_3$H$_{11}$ & a=2.891 \AA\ & Ge$_1$(4e) & 0.0000    & 0.0000 & 0.1750 \\ 
              & c=9.845 \AA\ & Ge$_2$(2b)& 0.0000    & 0.0000 & 0.5000 \\ 
              &              & H$_1$(8i) & 0.2248    & 0.0000 & 0.3320 \\
              &              & H$_2$(8i) & 0.7377    & 0.0000 & 0.0351 \\
              &              & H$_3$(4f) & 0.0000    & 0.5000 & 0.1031 \\
              &              & H$_4$(2c) & 0.0000    & 0.5000 & 0.2500 \\
              
 \hline
{\it C2/m\/}-GeH$_{4}$& a=10.226 \AA\     & Ge$_1$(2b) & 0.0000 & 0.5000 & 0.0000 \\
              & b=2.967 \AA\      & Ge$_2$(4i) & 0.8483 & 0.0000 & 0.6037 \\
              & c=2.922  \AA\     & H$_1$(8j)  & 0.3501 & 0.2383 & 0.1187 \\
        & $\beta$=74.46$^{\circ}$ & H$_2$(4i)  & 0.2822 & 0.0000 & 0.9765 \\
              &                   & H$_3$(4i)  & 0.2806 & 0.0000 & 0.6187 \\
              &                   & H$_4$(4i)  & 0.4274 & 0.0000 & 0.5731 \\
              &                   & H$_5$(4i)  & 0.9953 & 0.0000 & 0.7488 \\ 
 \hline
\end{tabular*}
 \label{tab-1}
\end{table}

\newpage

\begin{table}[ht]
\centering
\caption{\bf The calculated EPC parameter ($\lambda$), logarithmic average phonon frequency ($\omega_{log}$) and critical temperature ({\it T\/}$_c$) (with $\mu^*$ = 0.10 and 0.13) for 
{\it C2/m\/}-GeH$_{4}$ and {\it I$\bar{4}$m2\/}-Ge$_3$H$_{11}$ at given pressures.}
 \begin{tabular*}{9cm}{@{\extracolsep{\fill}}ccccc}
\Xhline{1pt} 
 Structure & Pressure (GPa) & $\lambda$ & $\omega_{log}$ (K) & {\it T\/}$_c$ (K) \\ [0.5ex] 
 \Xhline{1pt} 
  \multirow{4}{*}{{\it C2/m\/}-GeH$_{4}$} & \multirow{2}{*}{280} & \multirow{2}{*}{0.895}   & \multirow{2}{*}{1162}  & 67 ($\mu^*$=0.10)   \\ 
                      &   &   &   &  56 ($\mu^*$=0.13) \\

 \cline{2-5}
                  & \multirow{2}{*}{300}  & \multirow{2}{*}{0.867}    & \multirow{2}{*}{1154} & 63 ($\mu^*$=0.10) \\
                  &     &         &        & 52 ($\mu^*$=0.13) \\
\Xhline{1pt}

 \multirow{6}{*}{{\it I$\bar{4}$m2\/}-Ge$_3$H$_{11}$} & \multirow{2}{*}{285} & \multirow{2}{*}{0.721}    &  \multirow{2}{*}{1155} & 43 ($\mu^*$=0.10)\\ 
                  &     &       &    & 34 ($\mu^*$=0.13) \\
 \cline{2-5}

                   & \multirow{2}{*}{300}  & \multirow{2}{*}{0.690}    & \multirow{2}{*}{1140} & 38 ($\mu^*$=0.10) \\
                  &     &         &        & 29 ($\mu^*$=0.13) \\
  \cline{2-5}
                   & \multirow{2}{*}{320}  & \multirow{2}{*}{0.668}    & \multirow{2}{*}{1127} & 35 ($\mu^*$=0.10) \\
                  &     &         &        & 26 ($\mu^*$=0.13) \\
\Xhline{1pt} 
 
 \label{tab-2}
\end{tabular*}
\end{table}

\newpage
%\begin{figure*}[h]
%\centerline{\includegraphics[width=1.0\textwidth]{Convex.pdf}}
%\caption{Predicted formation enthalpy of  Ge$_{1-x}$H$_x$ as a function of H concentration at given pressures. Open circles above the convex hull show unstable compounds with respect to decomposition into the two adjacent phases on the convex hull, while solid circles show thermodynamically stable compounds. Pure Ge structures are consistent with Ref. \cite{PhysRevB.62.11388}, and pure H phases are taken from Ref. \cite{Pickard2007}.}
%\label{fig1}
%\end{figure*} 

%\begin{figure*}[h]
%\centerline{\includegraphics[width=0.5\textwidth]{Tc_vs_GPa.pdf}}
%\caption{Calculated critical temperatures T$_c$ for GeH$_4$ (star symbols) and Ge$_3$H$_{11}$ %(square symbols) as a function of pressure.} \label{Tc_vs_GPa}%!!!!
%\end{figure*}

\end{document}